\documentclass[prl,aps,showpacs,twocolumn,unsortedaddress]{revtex4}
\pdfoutput=1
\usepackage{graphics, bm}
\usepackage{psfrag}
\usepackage{amsmath}
\usepackage{amssymb}
\usepackage{epsfig}
\usepackage{subfigure}
\usepackage{grffile}
\newcommand{\eps}{\epsilon}
\newcommand{\beq}    {\begin{equation}}
\newcommand{\enq}    {\end{equation}}
\newcommand{\ceq}[1] {(\ref{#1})}

\newcommand{\rr}{{\bf r}}
\newcommand{\qq}{{\bf q}}

\newcommand{\nav}    {\langle n\rangle}
\newcommand{\nimp}   {n_{\rm imp}}
\newcommand{\Var}    {{\rm Var}}
\newcommand{\nrms}   {n^{(\rm rms)}}
\newcommand{\vrms}   {V_{D,{\rm sc}}^{(\rm rms)}}
\newcommand{\deltacr}{\Delta_{\rm cr}}
\newcommand{\rb}     {r_{\rm sc}}
\newcommand{\sio}    {${\rm SiO_2}$}
\begin{document}

\title{Inhomogenous electronic structure, transport gap, and percolation threshold in disordered bilayer graphene}

\author{E. Rossi$^1$, S. Das Sarma$^2$}
\affiliation{
             $^1$Department of Physics, College of William and Mary, Williamsburg, VA 23187, USA\\
             $^2$Condensed Matter Theory Center, Department of Physics,
             University of Maryland, College Park, MD 20742-4111, USA
            }
\date{\today}

   
\begin{abstract}
The inhomogenous real-space electronic structure of gapless and gapped disordered bilayer graphene is calculated 
in the presence of quenched charge impurities. 
For gapped bilayer graphene we find that for current experimental
conditions the amplitude of the fluctuations of the screened disorder potential is of the order of
(or often larger than)
the intrinsic gap, $\Delta$, induced by the application of a perpendicular electric field.
We calculate the crossover chemical potential, $\deltacr$, 
separating the insulating regime from a percolative regime in which
less than half of the area of the bilayer graphene sample is
insulating. 
We find that most
of the current experiments are in the percolative regime with
$\deltacr\ll\Delta$. The huge suppression of $\deltacr$
compared with $\Delta$ provides a possible explanation for the large
difference between 
the theoretical band gap $\Delta$
and the experimentally extracted transport gap.
\end{abstract}


\maketitle


One of the unique properties of single layer graphene, SLG, \cite{novoselov2004} is its 
high room-temperature electronic mobility \cite{dassarma2010rmp}. This fact
makes graphene of great interest for possible technological applications.
However, the lack of a band gap implies that in SLG the current can never
be turned off completely (i.e. SLG has a very low on/off ratio in the engineering jargon) 
and therefore limits the possible use
of SLG in transistor or switching applications. The mobility of bilayer graphene, BLG,
is normally lower than the mobility of SLG
\cite{morozov2008,xiao2010,hwang2009}, but it can be very high
when Boron-Nitride is used as a substrate
\cite{dean2010,dassarma2011}.
Most importantly, 
by applying a perpendicular electric field
\cite{castro2007, min2007, zhang-n-2009, oostinga2007, taychatanapat2010, zou2010, yan2010},
a gap of up to 250~meV can be opened in the band structure of BLG which should strongly enhance the on/off
switching ratio.
The strictly 2D nature of the carriers, the high room-temperature mobility, 
and the ability to open and tune $\Delta$, make BLG an extremely
interesting material
both from a fundamental and a technological point of view.
In recent BLG experiments \cite{oostinga2007, taychatanapat2010, zou2010} the activated transport gap 
has been found to be orders of magnitude smaller than $\Delta$.
This finding 
has been explained assuming that transport is in the variable
range hopping regime, VRH, 
\cite{oostinga2007, taychatanapat2010, zou2010},
or that edge modes might contribute significantly to transport
\cite{li2010}. 
Resonant scattering centers have been proposed as the dominant 
source of disorder \cite{wehling2010} in both SLG and BLG.
In gapped BLG the resonant scatterers would induce localized states
that would then mediate transport via VRH.
However scanning tunneling microscopy experiments have so far not
shown direct evidence of resonant states suggesting that their density
might be quite low, in addition 
no sign of localization is ever observed in ungapped BLG even in the presence of strong disorder.
On the other hand, the edge modes can significantly contribute only
if scattering between counter-propagating edge modes is suppressed. Because the
BLG band structure is characterized by two equivalent valleys
with opposite chirality, even small quantities of short-range defects
can mix the fermionic states of the two valleys
and greatly suppress the contribution of the edge modes to transport \cite{yan2010}.
These facts motivated us to look for a possible alternative explanation
to the smallness of the experimental BLG transport gap  compared to $\Delta$
based on the disorder-induced massive breakdown of momentum conservation.

Our model is based on the assumption that charge impurities are the dominant 
source of disorder in exfoliated BLG samples.
There is ample evidence \cite{dassarma2010rmp} that this assumption is at least consistent
with most of the transport experiments on gapless SLG and BLG although
other scattering sources might play an important role
\cite{wehling2010, katsnelson2008}.
In particular we assume the charge impurities
to be located at a typical distance $d\approx 1$~nm from the graphenic layer
and to be uncorrelated. In reality some degree of correlation
is expected but it does not affect qualitatively our results.
In the presence of charge impurities the carrier density becomes
strongly inhomogenous.
The importance of density inhomogeneities for the understanding of 
the physics, especially transport, of 2D electronic systems has been appreciated in the context
of the Quantum Hall, QH, effect \cite{disorderQH} and of {\em standard}
2D electron gases, 2DEGs, in which the gap between the hole-band and the electron-band is much larger
than the disorder strength \cite{disorderLargeGap}. 
The case of gapped BLG is different from these cases because
no magnetic field is present and so the dispersion is not broken-up
in Landau Levels and yet the band-gap is small compared to the strength of the disorder
potential. 
We emphasize, moreover, 
that the transport phase diagram, which is the main topic of our work , has never before been addressed in the literature for any system.
In BLG we therefore have the unique condition of a small band gap between
non-degenerate valence and conduction bands. 
The most striking
and counterintuitive consequence of this fact is that, as we show below, in gapped BLG an {\em increase}
of the disorder strength can drive the system from being an insulator to be a bad metal
when the chemical potential, $\mu$, is within the gap.
This behavior is the opposite of what happens in a
standard 2DEG in which an increase of the disorder strength drives the system 
to an insulating state. By showing that the disorder can effectively drive BLG to
a metal state even when $\Delta$ is finite and $\mu$ is well within
the gap, our work provides a compelling possible explanation for the large discrepancy
between $\Delta$ of gapped BLG and the gap extracted from transport measurements.
We define an effective {\em real-space} gap, $\deltacr$, that determines the transport
properties and show that for disorder strengths typical in current experiments it can
be zero even for $\Delta$ as large as 150~meV.
Although our specific calculations are carried out for gapped BLG
systems, the general idea developed here, namely a spatially
fluctuating local band gap in the presence of charged impurity
disorder induced inhomogeneity, should apply to other systems, and we
believe that the same idea could explain the experimental finding \cite{nanoribbons}
of rather small transport band gaps in graphene nanoribbon experiments
where percolation effects are known to be important.

Our main goal is to find a qualitative  explanation for the smallness
of the transport gap compared to $\Delta$ for the situation when charge impurities
are the dominant source of disorder. 
To calculate the electronic structure in the presence  of the 
disorder potential due to charge impurities we use the Thomas-Fermi theory, TFT.
For SLG the TFT results \cite{rossi2008} compare well with density functional theory, DFT,
results \cite{polini2008} and experiments \cite{martin2008, zhang2009,deshpande2009} 
as long as the impurity density $\nimp$ is not too low ($\gtrsim 10^{11}{\rm cm}^{-2}$) \cite{brey2009}.
These results suggest that TFT might give reasonable results also
for disordered BLG.
TFT is valid when the density profile, $n(\rr)$, satisfies the inequality $|\nabla n/n|<k_F$,
with $k_F=\sqrt{\pi n}$ the Fermi wavevector. As shown below the density varies on length
scales of the order of 10-20~nm whereas in the metallic regions $n\approx 2\times 10^{12}{\rm cm}^{-2}$, so that the
inequality  $|\nabla n/n|<k_F$ is only marginally satisfied. A complete quantitative validation
of the TFT results can only be achieved by comparison to DFT results that
however are not yet available for disordered BLG. On the other hand due to the
computational cost DFT cannot be used to calculate disorder averaged quantities
that are needed to extract the transport properties. TFT is therefore the only
approach that can be used to address the issue of transport in BLG in the presence
of long range disorder and, given that our goal is the {\em qualitative}
understanding of the large difference between transport gap and $\Delta$,
is also adequate. Moreover, the strong dependence of the transport gap
on the details of the experiments (like the temperature range) makes
a quantitative comparison of theory and experiments almost impossible.
For this reason, and given the limited quantitative accuracy of TFT, for 
the band structure of BLG we use the simple model of 2 parabolic
bands with effective mass $m^*=0.033 m_e$.
The simplicity of this model allows us to identify the few parameters that
affect the qualitative features of the results and makes 
our findings relevant also to standard parabolic 2DEGs.

%
The TFT energy functional is given by:
\begin{align}
 E[n] =&\int d\rr\left(\frac{\pi\hbar^2}{4m^*} n^2 + \frac{\Delta}{2}|n|\right) + 
       \frac{e^2}{2\eps}\int d\rr'\int d\rr \frac{n(\rr)n(\rr')}{|\rr-\rr'|}  \nonumber \\
       &+\frac{e^2}{\eps}\int d\rr V_D(\rr)n(\rr) -\mu\int d\rr n(\rr). 
 \label{eq:en}
\end{align}
where $\eps$ the static background dielectric constant,
and $eV_D/\epsilon$ the bare disorder potential.
The first term is the kinetic energy, the second the Hartree part of the 
Coulomb interaction, and the third is the contribution due to the disorder potential.
Assuming $\eps=4$ for the substrate, as appropriate for graphene on \sio, in 
the remainder we set $\eps=2.5$,
average of the dielectric constant of vacuum and substrate.
By differentiating $E[n]$ with respect to $n$ we find:
\begin{align}
 \frac{2m^*}{\hbar^2}\frac{\delta E}{\delta n} =& \pi n + \frac{m^*\Delta}{\hbar^2}{\rm Sign}(n) + 
                             \frac{1}{2\rb}\int\frac{n(\rr') d\rr'}{|\rr-\rr'|} \nonumber \\ 
                             &+\frac{V_D(\rr)}{\rb} -  \frac{2m^*\mu}{\hbar^2}.
\label{eq:deltaE}
\end{align}
where we have introduced the {\em screening length} $\rb \equiv [(2e^2m^*)/(\eps\hbar^2)]^{-1}\approx 2$~nm.

At the energy minimum, $\delta E/\delta n =0$.
For the gapless case this equation can be solved analytically in momentum space
to find
\beq
 n(\qq) = -\frac{V_D(\qq)}{\pi \rb}\frac{q}{q+\rb^{-1}}.
 \label{eq:nq}
\enq
Let
$\hat V_{D,{\rm sc}}(\rr)=\rb^{-1}V_D(\rr)+ (1/2\rb)\int d\rr' n(\rr')/|\rr-\rr'|$
be the screened disorder potential. Using Eq.~\ceq{eq:nq} we find:
\beq
 \hat V_{D,{\rm sc}}(\qq)= \frac{V_D(\qq)}{\rb} \frac{q}{q+\rb^{-1}}=-\pi n(q).
 \label{eq:vd}
\enq
We have $V_D(\qq)=A(\qq) e^{-qd}/q$, where $A(\qq)$
are random numbers 
with Gaussian distribution such that
$\langle A\rangle =0$ and $ \langle A^2\rangle = \nimp$,
where the angle brackets denote average over disorder realizations.
Using Eq.~\ceq{eq:nq},~\ceq{eq:vd} and the statistical properties
of $A(\qq)$, for the variance of the density, $\Var(n)$, we find
\beq
 \Var(n)=\frac{2\nimp}{\pi\rb^2}\int_0^\infty dq \frac{q e^{-2qd}}{(q+\rb^{-1})^2} = \frac{2\nimp}{\pi\rb^2} f(d/\rb)
\enq
with $f(d/\rb) = e^{2d/\rb}(1+2d/\rb)\Gamma(0,2d/\rb) - 1$,
a dimensionless function
(here $\Gamma(a,x)$ is the incomplete gamma function).
For small $d/\rb$, $f=-1-\gamma-\log(2d/\rb)+O(d/\rb)$ (where $\gamma=0.577216$ is the Euler constant), whereas for $d\gg \rb$
$f=1/(2d/\rb)^2 + O((d/\rb)^{-3})$.
For the root mean square of the density, $\nrms$, and screened disorder
potential, $\vrms=(\hbar^2/em^*)\hat V_{D,{\rm sc}}$, we then have:
\beq
 \nrms        = \frac{\sqrt{\nimp}}{\rb}\left[\frac{2}{\pi}f(d/\rb)\right]^{1/2}; \hspace{0.25cm}
 \vrms = \frac{\hbar^2\pi}{2m^*} \nrms.\
 \label{eq:nrms}
\enq
\begin{figure}[!!!t]
 \begin{center}
  \includegraphics[width=8.5cm]{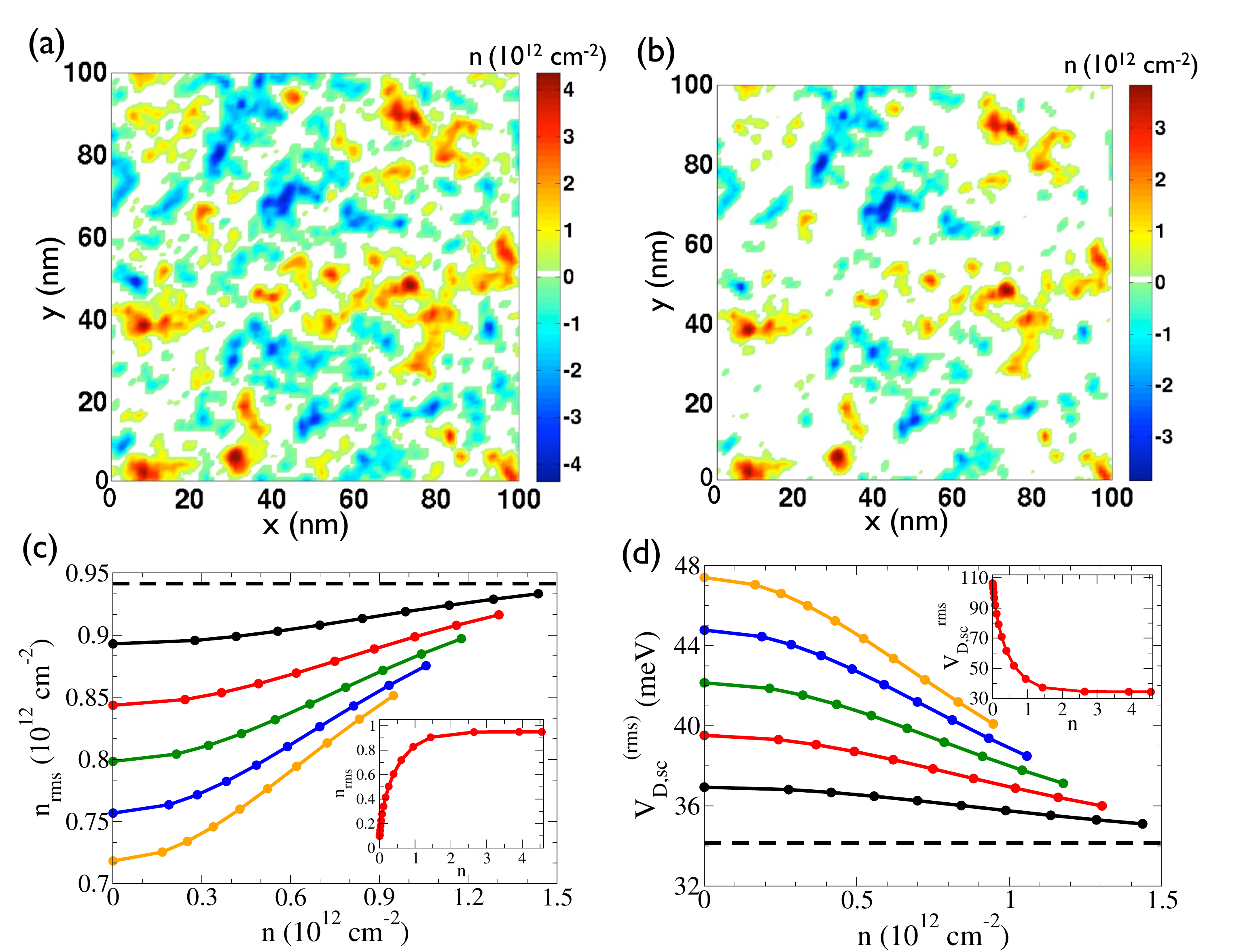}
  \caption{
           (Color online).
           Color plot of $n(\rr)$ at the CNP for a single disorder realization in gapped
           BLG, for 
	   $n_{imp}=8\times 10^{11}\;{\rm cm^{-2}}$, $d=1$~nm, and $\Delta=125$~meV (a) and $\Delta=250$~meV (b).
           The white areas in (a) and (b) represent insulating regions.
           $\nrms$ (c) and $\vrms$ (d) as a function of doping for $\nimp=3\times10^{11}$~cm$^{-2}$ and different values
           of $\Delta$, from top to bottom in (c) (bottom to top in (d))
           $\Delta=12, 23, 35, 46, 55$~meV. The dashed line in (c) and
           (d) shows the value of $\nrms$ and $\vrms$ respectively for
           the gapless case. The insets show the
           results for $\Delta=500$~meV \cite{note01}.
          } 
  \label{fig:gapped}
 \end{center}
\end{figure} 

In the presence of a band gap the equation $\delta E/\delta n=0$
becomes nonlinear and an analytic expression for $\nrms$ is not readily obtainable.
We have solved the problem numerically for $100\times 100$~nm samples,
with 1~nm spatial discretization, considering several (1000 or more) disorder realizations
to then calculate the disorder averaged quantities. 
In Fig.~\ref{fig:gapped}~(a) and (b) we show our calculated carrier density
landscapes for single disorder realizations for two values of $\Delta$,
both with the same $\nimp$ and $\mu$
fixed at the charge neutrality point, CNP.
In the absence of disorder both situations in Fig.~\ref{fig:gapped}~(a) and (b)
will manifest zero carrier
density throughout with a pure intrinsic band gap at all spatial
points (i.e. both sets of plots will be completely 'white' in color
since we are explicitly at T=0 with no thermal inter-band
excitations).  
Fig.~\ref{fig:gapped}~(c) and (d) show the dependence of $\nrms$ and $\vrms$, respectively, on
the doping $n$ for values of the gap between 12 and 55 meV. 
$\nrms$ is suppressed
close to the CNP, due to the large area covered by insulating regions ($n=0$) whereas
$\vrms$ is higher close to the CNP due to the lack of screening.
The results of Fig.~\ref{fig:gapped}~(c) and (d) are well fitted by the
scalings 
$\nrms = \nrms(\Delta=0)-9\times 10^{-3}\Delta^{0.78} e^{-n^2/b_n}$, 
$\vrms = \vrms(\Delta=0)+0.24\Delta e^{-n^2/b_v}$, with 
$b_n=1.83 e^{-\Delta/98.5}$,
$b_v=2.31 e^{-\Delta/73.26}$
and $n$ and $\Delta$ expressed in units of $10^{12}{\rm cm}^{-2}$ and meV respectively.

\begin{figure}[!!!t]
 \begin{center}
  \centering
  \includegraphics[width=8.5cm]{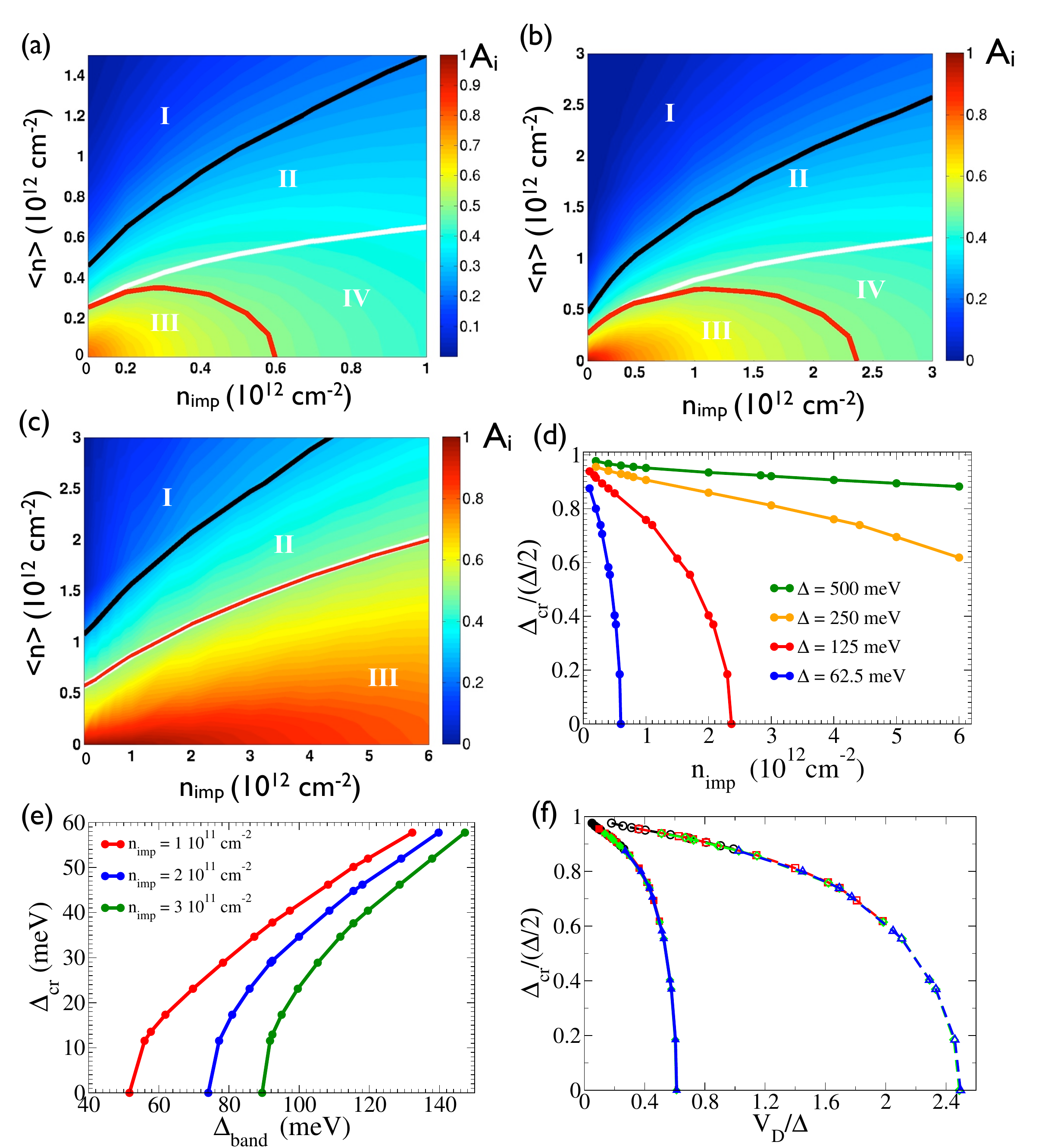}
  \caption{
           (Color online). 
           Color plots showing $A_i$ as a function of $\nimp$ and $\nav$
           for $\Delta=125$~meV, (a), $\Delta=250$~meV, (b), and $\Delta=500$~meV, (c).
           (d) $\deltacr$ as a function
           of $\nimp$ for different values of $\Delta$.
           (e) $\deltacr$ as function of $\Delta$ for different values of $\nimp$.
           (f) $\deltacr/(\Delta/2)$ as a function of $V_D/\Delta$. The filled (unfilled) symbols, connected by the solid (dashed) line,
           show the dependence of $\deltacr/(\Delta/2)$  with respect to 
           $\vrms/\Delta$ ($(eV_D^{\rm (rms)}/\epsilon)/\Delta$). 
           The different symbols, circles, squares, diamonds, triangles, show the results for
           $\Delta=500, 250, 125, 62.5$~meV respectively.
           \cite{note01}.
         } 
  \label{fig:gblg}
 \end{center}
\end{figure}

The analysis of the disorder averaged results obtained from the TFT
can be used to qualitatively understand the electronic transport
in the highly inhomogeneous density landscape of gapped BLG. 
Let $A_i$ be the disorder averaged fraction of the area occupied
by the insulating regions (white in Figs.~\ref{fig:gapped}~(a) and (b)).
The color plots in Fig.~\ref{fig:gblg}~(a)-(c) show the dependence of
$A_i$ on $\nimp$ and the doping $\nav$ for three different values of $\Delta$.
At high dopings and low disorder $A_i$ is fairly small.
As $\nav$ decreases some hole puddles and insulating regions start appearing.
The black line identifies the contour for which $\nrms=\nav$.
Below the black line $\nrms >\nav$ and we are in the strongly
inhomogenous regime. The white line identifies the contour for which
the area fraction occupied by electron puddles is equal to 50\%.
Below this line more than half of the sample area is occupied by
hole puddles and insulating regions. Finally, the red line shows
the contour for which is $A_i=0.5$. Below this line more than
half of the sample is covered by insulating regions.
The counterintuitive result is that 
close to the CNP $A_i$ becomes smaller as $\nimp$ increases with 
$\nav$ fixed. 
This is a consequence of the fact that
as $\nimp$ increases, the disorder becomes strong enough
to bring the Fermi level inside the conduction or the valence band.
Using the contour lines overlaid on the color-plot for $A_i$ we
can qualitatively identify different semiclassical transport regimes.
{\em Regime I}: Weak disorder where $\nrms\ll \nav$.  The system is a good
metal with an almost uniform density landscape.
{\em Regime II}: Strong disorder $\nrms > \nav$. In this regime the density
landscape breaks up in puddles and insulating regions. 
Because the area fraction covered by electrons is larger than 1/2
the system behaves like a metal via percolation through the electron puddles.
{\em Regime III}: Weak-moderate disorder, $|\mu|<\Delta/2$, $\vrms<\Delta/2$.
Most of the sample area is insulating and the system is an insulator.
{\em Regime IV}: In this regime the disorder is so strong, $\vrms\gg\Delta/2>|\mu|$, that, despite the finite
gap, $A_i$ is less than 50\%
but neither the electron puddles, nor the hole puddles alone cover more than 50\%
of the total area. The system should behave as a bad metal with the conductance
determined by tunneling events across few narrow insulating regions separating the 
electron-hole puddles similar to transport at the CNP in gapless BLG
\cite{hwang2010} and SLG \cite{rossi2009}.
It is obvious, looking at the color plots in Fig. 2(a)-(c) that at
lower (higher) values of $\Delta$ (disorder) the apparent transport
gap will be strongly suppressed and in fact will be vanishingly
small.  We believe that this is the current physical situation in
existing BLG systems.  

To identify the transport regime is useful to introduce
the critical value of chemical potential, $\deltacr$, for which $A_i$=50\%.
For $\mu>\deltacr$ ($\mu<\deltacr)$ the system is expected to behave as bad metal (an insulator).
For $\Delta\gg \vrms$, $\deltacr\approx\Delta/2$ is almost independent
of the disorder strength. 
For small gaps, $\Delta\lesssim \vrms$, $\deltacr$
depends strongly on the disorder strength, and $\deltacr\ll\Delta$. This is shown 
in Fig.~\ref{fig:gblg}~(d) in which the calculated $\deltacr$ as a function of
$\nimp$ is shown for four different values of $\Delta$.
For fixed $\Delta$, $\deltacr$ decreases as $\nimp$ increases.
We see that for $\Delta\lesssim 150$~meV, for impurity
densities of the order of the ones estimated in current experiments on
exfoliated BLG ($\nimp\approx 10^{12}\;{\rm cm}^{-2}$), $\deltacr$ 
can be orders of magnitude smaller than $\Delta$.
Fig.~\ref{fig:gblg}~(e) shows $\deltacr$ as a function of $\Delta$
for different values of $\nimp$. We see that for $\Delta\approx 100$~meV
$\deltacr\approx 0$ already for $\nimp\approx  3\;10^{11}{\rm cm}^{-2}$.
In the presence of spatial correlations among impurities
the density inhomogeneities are expected to be reduced and therefore
the value of $\nimp$ for which $\deltacr\to 0$ would increase.
Finally Fig.~\ref{fig:gblg}~(f) shows the scaling of $\deltacr$
on the strength of the disorder potential, $V_D$.
The points connected
by solid (dashed) lines show the dependence of $\deltacr/(\Delta/2)$ with respect
to $\vrms/\Delta$ ($(eV_D^{\rm (rms)}/\epsilon)/\Delta$), with $V_D^{\rm (rms)}$
the rms of the bare disorder potential.
In both cases, by normalizing
both $\deltacr$ and $V_D$ with the band-gap, we find that the results
obtained for different values of $\Delta$ collapse on a single curve
that does not depend on $\Delta$ and away from $V_D=0$ scales approximately
as
$1-ae^{b V_D/\Delta}$, with $(a=0.02, b=1.52)$
for $V_D=eV_D^{\rm (rms)}/\epsilon$,
and $(a=0.015, b=6.65)$ for $V_D=\vrms$.

In summary, using a simple 2-bands model and TFT, we have characterized the density inhomogeneities
in gapless and gapped BLG. For gapless BLG we have found analytic 
expressions for  $\nrms$ and $\vrms$ as a function of the experimental
parameters and shown that they do not depend on the doping and scale like
$\sqrt\nimp f^{1/2}(d/\rb)/\rb$.
For gapped BLG $\nrms$ ($\vrms$) is reduced (enhanced)
with respect to the gapless case, in particular in the vicinity of the CNP.
By calculating the disorder averaged fraction of
the sample area, $A_i$, covered by insulating regions we have qualitatively identified
four different transport regimes. We have shown that most of the current
experiments are expected to be in  a regime, regime {\em IV}, in which the
disorder is strong enough to reduce $A_i$ below 50\% even at zero doping.
In this regime, gapped BLG is expected to behave like a bad metal in which
transport is dominated by hopping processes between electron and hole
puddles that cover most of the sample. 
The value of the chemical potential
$\deltacr$
for which $A_i=0.5$ identifies the crossover between the insulating
regime and regime {\em IV}. We have shown how $\deltacr$ depends
on the impurity density $\nimp$, the strength of the screened disorder potential,
and the theoretical band gap $\Delta$. 
We believe the reduction of $\deltacr$ as a function
of $\nimp$ is the qualitative resolution of the
contradiction between $\Delta$ and the
transport gap.
A clear prediction of our theory is that in cleaner BLG (e.g. on boron
nitride substrates) there should be close agreement between the
transport gap and $\Delta$.
We also predict agreement between $\Delta$ and the
transport gap for very large values of $\Delta$.

This work is supported by US-ONR and NRI-SWAN. E.R. acknowledges
support from the Jeffress Memorial Trust, Grant No. J-1033. 
Computations were carried out on the 
University of Maryland High Performance Computing Cluster (HPCC)
and the SciClone Cluster at the College of William and Mary.



\begin{thebibliography}{28}
\bibitem[{not()}]{note01}
\bibinfo{note}{Even for large electric fields in BLG $\Delta$ saturates to
  $\sim250$~meV. Results for $\Delta>250$~meV are shown to illustrate
  qualitatively the situation $\Delta\gg\vrms$ typical in standard parabolic
  2DEGs such as the ones created in GaAs and Si quantum wells}.
\end{thebibliography}

\end{document}